\documentclass[aps,prb,reprint,superscriptaddress,showpacs]{revtex4-1}

\usepackage{graphics} 
\usepackage{amsmath} 
\usepackage{graphicx} 
\usepackage{dcolumn} 
\usepackage{bm} 
\usepackage{color}
\usepackage{subfigure} 
\usepackage{bbm, dsfont,url}

\newcommand{\eps}{\varepsilon}
\newcommand{\bd}{\hat{b}^{\dag}}

\begin{document}

\title{Quench dynamics of a disordered array of dissipative coupled
  cavities}

\author{C. Creatore}
\affiliation{Cavendish Laboratory, University of Cambridge, CB3 0HE Cambridge, United Kingdom}
\author{R. Fazio}
\affiliation{NEST, Scuola Normale Superiore and Istituto Nanoscienze-CNR, I-56127 Pisa, Italy}
\author{J. Keeling}
\affiliation{Scottish Universities Physics Alliance, School of Physics and Astronomy, University of St Andrews, St Andrews KY16 9SS, United Kingdom}
\author{H. E. T\"{u}reci}
\affiliation{Department of Electrical Engineering, Princeton University, Princeton, New Jersey 08544, USA}


%


\begin{abstract}
  We investigate the mean-field dynamics of a system of interacting
  photons in an array of coupled cavities in presence of dissipation
  and disorder. We follow the evolution of on an initially prepared
  Fock state, and show how the interplay between dissipation and
  disorder affects the coherence properties of the cavity emission and
  that these properties can be used as signatures of the many-body phase of the
  whole array.
\end{abstract}

\maketitle

\section{Introduction}

The idea of understanding the behaviour of complex quantum many-body
systems using experimentally controllable {\em quantum} {\em
  simulators} can be traced back to a pioneering keynote speech given
by Richard Feynman in 1982~\cite{Feynman1982}.  After thirty years,
quantum simulation is now a thriving field of
research~\cite{Cirac2003,Nori2009}, driven by the increasing ability
to design and fabricate controllable quantum systems, in contexts
ranging from superconducting-circuits~\cite{Houck2012a} to ultracold
atoms~\cite{Bloch08} or trapped ions~\cite{monroe13}.  These
systems allow the realisation of archetypal models and the 
exploration of new physical regimes. An area of recently developing
interest has been coupled cavity arrays, lattices of coupled
matter-light systems, providing highly tunable but dissipative quantum
systems~\cite{hartmann2006,Greentree2006,Angelakis2007a,hartmann2008,fazio2010}.

Physical realisations of coupled cavity arrays have been proposed in a
variety of systems, such as photonic crystal
nanocavities~\cite{Vuckovic2012}, coupled optical
waveguides~\cite{Leppert13}, or lattices of superconducting resonators
operating in the microwave regime~\cite{Houck2012a,Houck2012b,Koch2013}.
While experiments have not yet probed the collective behaviour
predicted in large scale arrays, progress towards such realisations is
very encouraging.  For these different systems, the radiation modes
involved range from microwave to optical frequency; we will
nonetheless refer to these as coupled ``matter-light'' systems in the
following, taking ``light'' to refer to the radiation modes.

\begin{figure}[ht!]
\centering
\includegraphics[width=0.45\textwidth]{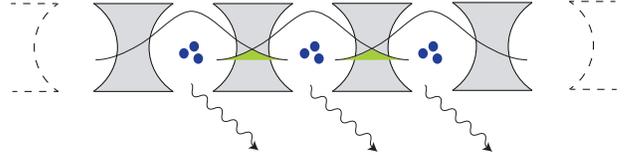} 
\caption{A sketch of a one-dimensional cavity array as described in the text.}
\label{fig:sketch}
\end{figure}

A generic coupled cavity array (CCA) consists of a lattice of
cavities, each supporting a confined photon quasi-mode. We refer to
quasi-modes~\cite{scully97} since the finite quality of the cavities
implies there will be coupling to the outside world.  As shown in
Fig.~\ref{fig:sketch}, the cavities are coupled through photon
hopping.  A purely optical system would be entirely linear, and thus
unable to simulate interacting many-body quantum systems. To introduce
nonlinearity requires coupling to matter (e.g. suitable optical
emitters such as semiconductor quantum dots or superconducting qubits,
indicated by blue circles in Fig.~\ref{fig:sketch}).  This leads to a
system of photons hopping on a lattice, with an on-site nonlinearity
and on-site losses.  

A wide variety of different microscopic models can be realised depending on the
precise design of the array~\cite{hartmann2008,fazio2010,Koch2013}. 
We consider here the archetypal case where the CCA can be mapped~\cite{hartmann2008} 
onto the Bose-Hubbard model~\cite{Fisher1989}.  This corresponds to the
regime of weak matter-light coupling, i.e. strong detuning between cavity
mode and matter degrees of freedom~\cite{Angelakis2012}.

In the absence of dissipation, the ground state phase diagram of the Bose-Hubbard model 
has been extensively studied~\cite{Bloch08}: At zero temperature a 
quantum phase transition results from the competition between on-site 
nonlinearities and inter-site photon hopping.  When the nonlinearities 
are strong and prevail over the hopping, the photons are localised by 
interactions, leading to an insulating Mott phase; in the opposite 
regime, when photon hopping dominates, a superfluid phase 
characterised by long-range coherence occurs.  Coupled matter-light 
systems however are naturally studied under non-equilibrium conditions, as 
there are invariably photon losses.  As such, a steady state in a 
coupled cavity array requires external pumping. The steady-state of 
cavity arrays, resulting from the competition of external driving and 
dissipation, has been the subject of much recent theoretical 
interest~\cite{Carusotto2009,Hartmann2010,Nissen2012,Angelakis2012,Fazio2013,Keeling2013}.

Besides the steady state, the transient dynamics of cavity arrays may
bring additional interesting information.  By engineering suitable
laser pulse sequences~\cite{Gross2007,Brierley2012} it is possible to
prepare a specific initial state and follow its susbequent evolution
by analysing the properties of the light escaping from the system.
This is the situation considered in~[\onlinecite{Tomadin2010}], considering
the evolution following preparation of a Fock state. Such a protocol
is equivalent to studying a quantum quench in an open system.  Quantum
quenches in closed systems are an intensively studied paradigm of
non-equilibrium dynamics of many-body systems~\cite{Polkovnikov2013},
and in the closed Bose-Hubbard model have been already addressed both
theoretically and experimentally,
see ~[\onlinecite{Polkovnikov2013,Kollath2007,Sciolla2010,Cheneau2012}]. Cavity
arrays appear to be ideally suited to explore the transient dynamics following
a quench.

Remarkably, even in the lossy system, the dynamics following such a
quench clearly map out a superfluid--insulator
transition~\cite{Tomadin2010}.  In fact, in mean-field theory these
can be directly related to the equilibrium phase
boundary~\cite{Tomadin2009,Tomadin2010}.  In particular, by rescaling
correlation functions by the decaying density one finds that the
behaviour at long times is distinct for values of hopping in the
superfluid and insulating phases --- the rescaled coherence vanishes
in the insulating phase, but attains a non-zero asymptote in the
superfluid phase.  As discussed further below, this can be traced back
to the way in which for the closed system quench, the linear stability
of an initial Mott state reproduces the equilibrium phase
diagram~\cite{Altman2002}.  Given the notable ability of the rescaled
correlations of the open system to reproduce the equilibrium phase
boundary, and the significant interest in the disordered Bose-Hubbard
model~\cite{Fisher1989,Gurarie2009,Bissbort2010,Pollet2013}, a natural question to ask is how the open
system quench dynamics are affected by disorder.  This is the question we
being to address in this paper.  Building on the results of
Tomadin~\textit{et al.}~\cite{Tomadin2010}, we analyse the
non-equilibrium mean-field dynamics of an array of non-linear coupled
cavities in presence of photon leakage and disorder in the on-site
cavity energies.

The paper is organised as follows: In Section 2 we introduce the model
used to describe the dissipative and disordered cavity array; in
Section 3 we analyse the dynamics of the correlation functions of the
cavity array, first summarising the results for the ideal
\textit{clean case} and then assuming a Gaussian distribution for the
on-site cavity energies, and compare the results; in Section 4 we briefly discuss 
an anomalous behaviour of the second order correlation function in presence of 
disorder; in Section 5 we summarise our conclusions.
\section{Model system and initial state}

As discussed above, we consider the Bose-Hubbard
model~\cite{Fisher1989} described by the Hamiltonian:
\begin{align}
  \hat{H}=\sum_{i}\frac{U}{2}\hat{n}_{i}(\hat{n}_{i}-1) + \eps_{i}\hat{n}_{i}-
  J\sum_{\langle i\, j\rangle}\bd_{i}\hat{b}_{j}.
\label{eq:hamil1}
\end{align}
Here $\bd_{i}$ ($\hat{b}_{i}$) creates (annihilates) a photon in the $i$-th
site and the corresponding number operator is $\hat{n}_i = \bd_i \hat{b}_i$.  The energy of the photon mode in the $i$th cavity is $\varepsilon_i$, $J$ denotes the
amplitude for inter-site photon hopping, while $U$ represents the
on-site nonlinearity resulting from the coupling to matter.  Including
also the presence of Markovian photon loss leads to the master equation:
\begin{equation}
  \label{eq:1}
  \partial_t \rho(t) = -i[\hat{H}, \rho(t)] + \kappa \sum_{i}
  \mathcal{D}[\hat{b}_i,\rho],
\end{equation}
where loss is described by the Lindblad term $\mathcal{D}[X,\rho]=2 X
\rho X^\dagger - X^\dagger X \rho - \rho X^\dagger X$.  The photon
lifetime is $(2\kappa)^{-1}$.  We consider the case where $U, \kappa,
J$ are independent of site, but there may be disorder in the energies
$\varepsilon_i$.  We will discuss below  the ``clean'' case in which
all $\varepsilon_i$ are the same, and the disordered case where we
will choose cavity energies $\varepsilon_i$ to be drawn from Gaussian
distribution having mean value $\bar{\varepsilon}=\eps_{0}$ and
variance $\sigma^{2}$, with $\sigma$ representing the strength of the
disorder.  The mean value $\bar\varepsilon$ can be removed by a gauge
transformation, and so may be set to zero without loss of generality. 
The numerical integration of the master equation has been implemented 
using a fourth- and fifth-order Runge-Kutta algorithm and we truncate the 
Fock basis for each cavity to $\{|n \rangle_{i}\}_{n=0}^{n_{\textrm{max}}}$ with 
$n_{\textrm{max}}=4$.

We explore the quench dynamics of this model in the mean-field
approximation $\rho=\prod_i \rho_i$, which decouples the photon
hopping between neighbouring cavities but allows for a spatially inhomogeneous solution. Such an approximation is expected to be valid
in the limit of large coordination $z$, where $z$ is the number of
nearest neighbours of each site. The dynamics of each cavity is
the governed by the master equation
\begin{align}
  &\partial_ {t}\rho_{i}(t)=\mathcal{L}\rho_{i}(t)\,,
\label{eq:master}  
  \\
    &\mathcal{L}=-i\left[\hat{h}_{i},\rho_{i}(t)\right] + \kappa \mathcal{D}[\hat{b}_{i},\rho_{i}(t)] \,,
\label{eq:liouvillian}
\\
&\hat{h}_{i}=\frac{U}{2}\hat{n}_{i}(\hat{n}_{i}-1) + \eps_{i}\hat{n}_{i}- J(\phi_{i}(t)\bd_{i} + \phi_{i}^{*}(t)\hat{b}_{i}) \,,
\label{eq:hamil3}     
\end{align}
where $\phi_{i}(t)=\sum_{j \in \text{nn}(i)}
\textrm{Tr}[\hat{b}_{j}\rho_j(t)]$ is summed over the $z$ nearest neighbours
of site $i$.  In the clean case all sites are equivalent and so
$\phi_i(t)=z \textrm{Tr}[\hat{b}_{i}\rho_i(t)]$, and the dimensionality only
enters via this factor $z$.  In the disordered case, each site evolves
separately, and the connectivity of the lattice does affect the
dynamics.

Our goal in the following discussion is to study the non-equilibrium dynamics following preparation of the cavity array in a product of Fock states, i.e. $\rho(0) = \prod_i
\rho_i(0), \rho_i(0)=|n_0 \rangle \langle n_0 |$. Because $\textrm{Tr}[\hat{b}_{i}\rho_i(0)]=0$ is a fixed point of the mean field equations, we consider a small deviation
away from such a Fock state, and instead consider $\rho_i(0) =
|\Psi_0\rangle \langle \Psi_0|$, where $|\Psi_0\rangle =
\sqrt{1-\eta^2} |n_0\rangle + \eta |n_0 -1\rangle$.  As discussed below, depending on the parameters, the Fock state may be either a stable or unstable fixed point, and if unstable, a small initial perturbation $\eta\ll 1$ will grow, and drive the array to a different asymptotic state.

\section{Dynamics of correlation functions}

\subsection{Clean case}
\label{sec:clean-case}

Before exploring the role played by disorder, we first summarise the
results of the clean case~\cite{Tomadin2010} and present also a 
discussion of the initial instability, first considered in~[\onlinecite{Tomadin2009}] 
and here thoroughly discussed. In the absence
of dissipation, it was shown~\cite{Altman2002} that when
$zJ/U>zJ/U|_{\textrm{cr}}$, where $zJ/U|_{\textrm{cr}}$ is the critical
value corresponding to the equilibrium superfluid--insulator phase
transition, an initial Fock state is linearly unstable.  As such, the
existence of this instability can be used to trace the equilibrium
transition between the Mott and the superfluid
phase~\cite{Tomadin2009,Altman2002} and, within our mean-field
analysis, a transition survives also in the presence of dissipation.

\begin{figure}[ht!]
\centering
\includegraphics[width=0.45\textwidth]{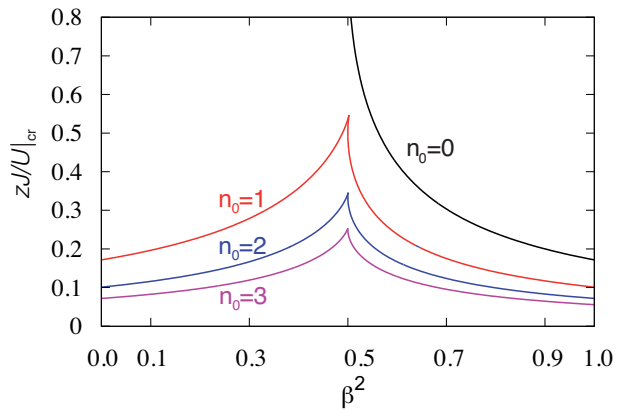} 
\caption{$\left.zJ/U\right|_\text{cr}$ as a function of $\beta^{2}$, the initial state being the statistical admixture 
$\rho(0)=\alpha^{2}|n_{0}\rangle\langle n_{0}| + \beta^{2}|n_{0}+1\rangle \langle n_{0}+1|$, with $\alpha^{2}+\beta^{2}=1$.}
\label{fig:jcritic}
\end{figure}

To see this instability for the clean case, one may write coupled
equations for $\rho_{n_0-1, n_0}, \rho_{n_0,n_0+1}$ (where $n_0$ is the 
occupation of the initial state, under the approximation
$\rho_{n_0,n_0} \simeq 1$, and that all other elements of $\rho$ are
negligible.  Denoting $X=(\rho_{n_0-1, n_0}, \rho_{n_0,n_0+1})^T$, one
finds that X obeys the equation $\partial_t X = M X$ with
\begin{widetext}
 \begin{equation}
   \label{eq:2}
   M=
   \left(
     \begin{array}{cc}
       iU(n_0-1) + i zJ n_0  -  \kappa(2n_0-1) & (i zJ + 2\kappa) \sqrt{n_0(n_0+1)} \\
       -i zJ \sqrt{n_0(n_0+1)}  & i U n_0  -i zJ (n_0+1) - \kappa (2n_0+1)
     \end{array}
   \right).
\end{equation}
\end{widetext}
Instability occurs when the real part of the eigenvalues of $M$ become
positive, as this corresponds to exponential growth of fluctuations. 
In the absence of dissipation ($\kappa=0$), the eigenvalues $\xi$ are given by:
\begin{equation}
   \xi = iU\left(n_0-\frac{1}{2}\right) - \frac{izJ}{2}
    \pm i \sqrt{4n_{0}zJU-\left(U-zJ\right)^{2}}
\end{equation}
The stability boundary is at the point where the eigenvalues
are both pure imaginary, 
and writing $\xi=i\mu$, one may show that $\text{Det}(i\mu\mathbbm{1}-M) =0$ is equivalent to:
\begin{equation}
  \label{eq:3}
  \frac{1}{zJ} = \frac{n_0}{\mu-U(n_0-1)} - \frac{n_0+1}{\mu- Un_0}
\end{equation}
which is the equilibrium phase boundary for the $n_0$th Mott lobe.  As
$zJ$ increases, the critical values of $\mu$ in the equilibrium phase
boundary approach each other, and at large enough $zJ$, there is no
longer any real value $\mu$ that can satisfy the above equation.  As
such, the instability of the Fock state corresponds to the locations
of the tips of the equilibrium Mott lobes,
$\left.zJ/U\right|_\text{cr} =(\sqrt{n_0+1} - \sqrt{n_0})^2$.  A comprehensive
discussion of the phase boundary at finite $\kappa$ can be found in~[\onlinecite{Tomadin2009}]. 
Further, this analysis allows one to study how $\left.zJ/U\right|_\text{cr}$ changes when the initial state is a statistical admixture, 
e.g. when $\rho(0)=\alpha^{2}|n_{0}\rangle\langle n_{0}| + \beta^{2}|n_{0}+1\rangle \langle n_{0}+1|$ ($\alpha^{2}+\beta^{2}=1$). 
In this case,  for $\kappa=0$, the eigenvalues $\xi$ are given by the solutions of:

\begin{eqnarray}
 \label{eq:mix}
  \frac{1}{zJ} =&-&\frac{i\alpha^{2}n_0}{\xi+iU(n_0-1)} + \frac{i(\alpha^{2}-\beta^{2})(n_0+1)}{\xi+iUn_0}\\ \nonumber
  &+& \frac{i\beta^{2}(n_0+2)}{\xi+iU(n_0+1)}\,.
\end{eqnarray}
We note that the ground state for any incommensurate filling would always be a superfluid phase.  
As such, in the equilibrium phase diagram, the critical hopping jumps
discontinuously as one varies the filling. Equation (\ref{eq:mix}) describes a
different question however; namely whether the critical hopping for a
\emph{linear} instability of the normal state evolves continuously.  This is answered by Fig.~\ref{fig:jcritic}, showing that 
the critical hopping $\left.zJ/U\right|_\text{cr}$ for this instability [obtained solving Eq.~(\ref{eq:mix}) for different $n_{0}$] 
does evolve continuously as a function of $\beta^{2}$. In the remainder of this manuscript we restrict to the case $\beta=0$. 
\begin{figure}[ht!]
  \begin{center}$
    \begin{array}{c}
      \includegraphics[width=0.45\textwidth]{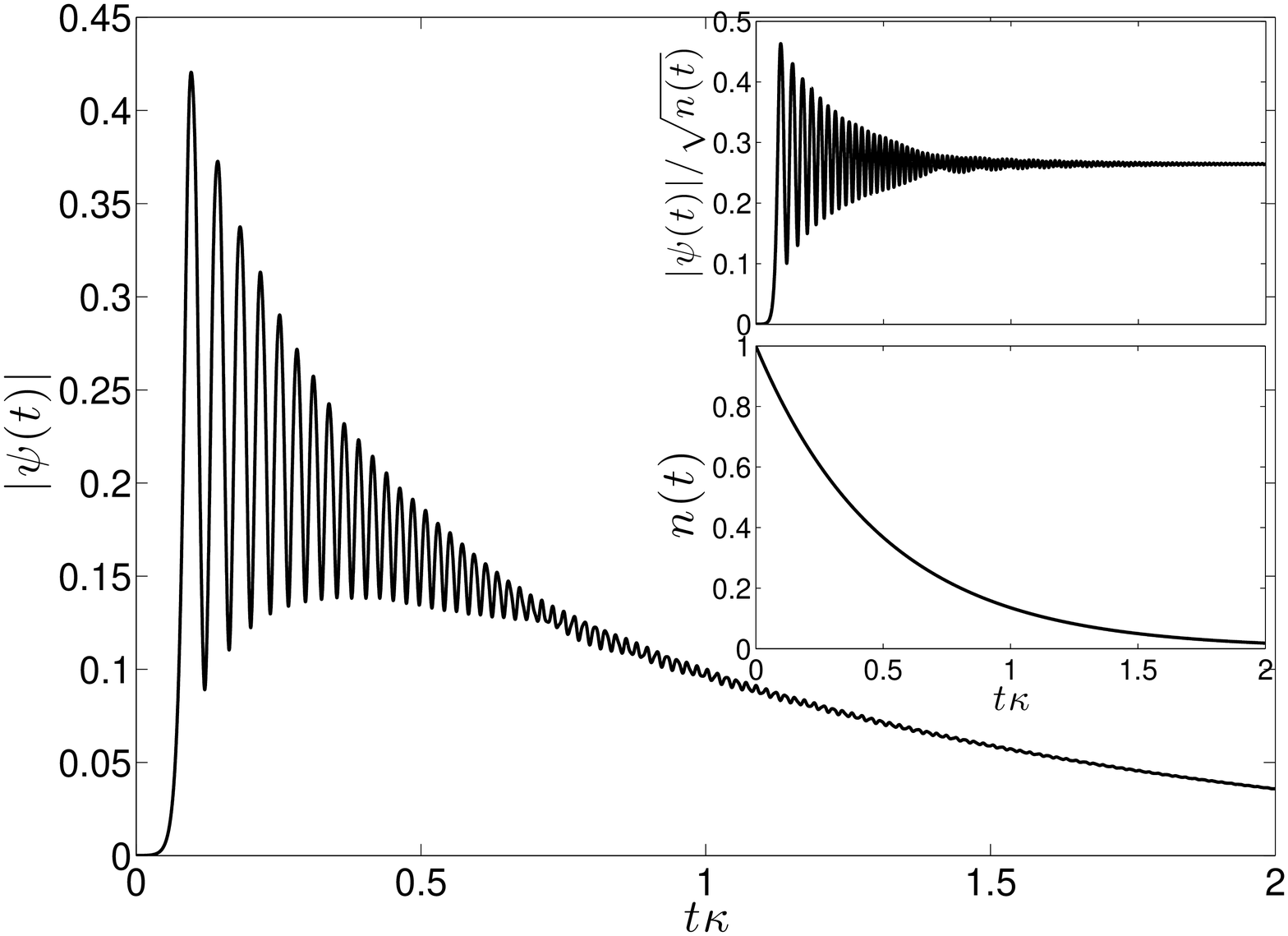} \\ 
      \includegraphics[width=0.45\textwidth]{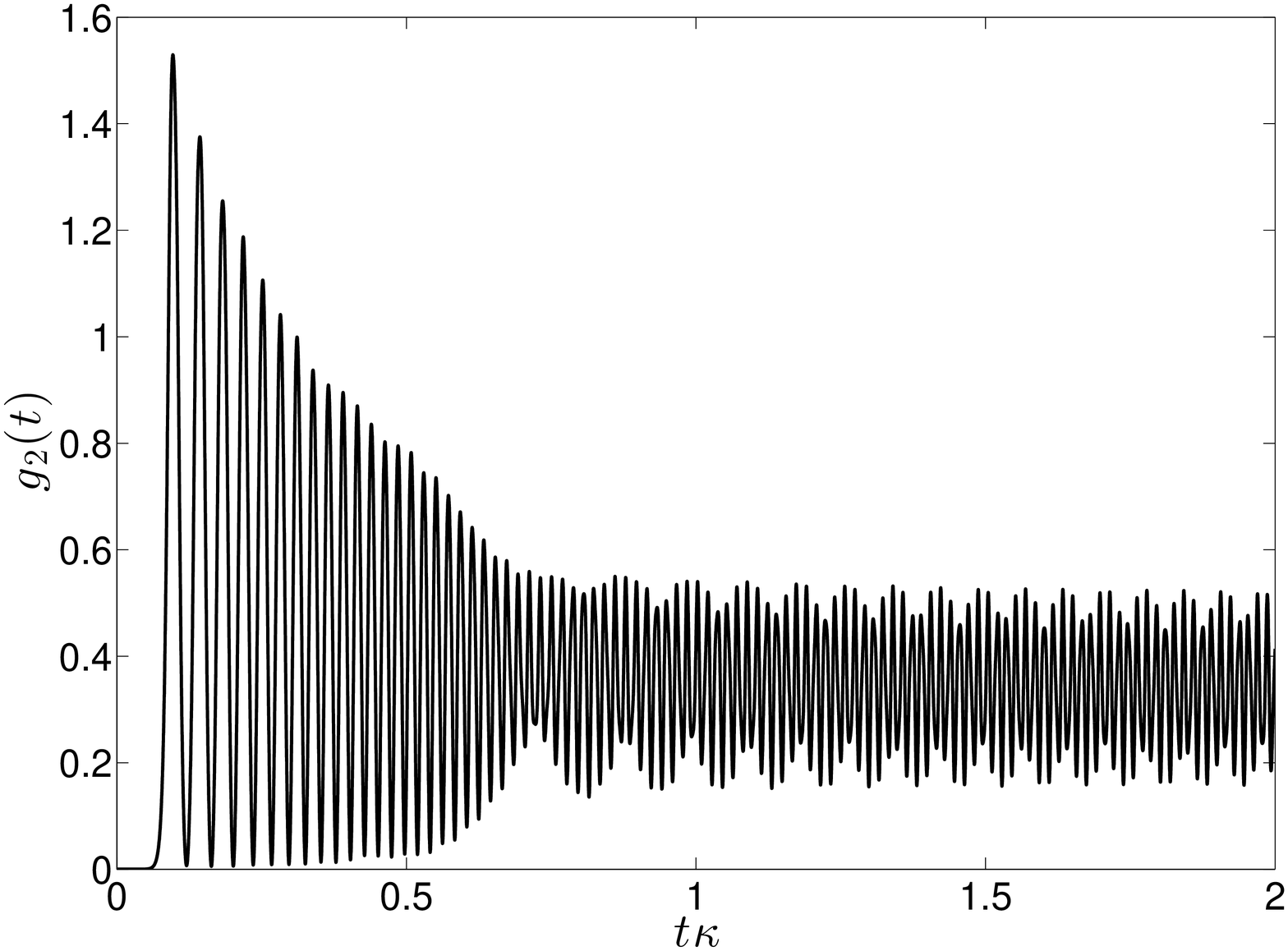}     
    \end{array}$
  \end{center}
  \caption{Dynamics of a one-dimensional array ($z=2$) of $N=10$
    coupled cavities in presence of a weak dissipation
    $\kappa=10^{-2}U$ and for $zJ=3U$.  Top panel: Time evolution of
    the absolute value of the order parameter $|\psi(t)|$.  For the
    same parameters of the main panel, the bottom inset shows the
    evolution of the average filling $\langle n\rangle=n(t)$, while
    the top inset shows the rescaled order parameter
    $\overline{\psi}(t)=|\psi(t)|/\sqrt{n(t)}$. Bottom panel: Time
    evolution of the zero-time delay second order correlation function
    $g_{2}(t)$.}
\label{fig:clean1}
\end{figure}

The consequence of this instability can be seen in the time evolution
of the coherence $\psi(t) = \textrm{Tr}[\hat{b}_i \rho_i(t)]$.
Figure~\ref{fig:clean1} shows the evolution of $\psi_i(t)$ for a clean
system with $zJ/U > \left.zJ/U\right|_{\text{cr}}$.  The figure is
plotted for parameters $\kappa=10^{-2}U$ and $zJ=0.3U$, with an
initial state with $n_0=1, \eta=10^{-5}$ --- we consider this same initial
state throughout the remainder of the manuscript.  In the initial time
range $t\kappa\lesssim\,0.1$, due to the instability, $\psi(t)$ grows 
exponentially even though the photon population $n(t) =
\textrm{Tr}[\hat{n}_i \rho_i(t)] $ decays exponentially
$n(t)=n(0)e^{-2\kappa\,t}$, see the bottom inset in left main panel of
Fig~\ref{fig:clean1}.  This exponential growth leads to a regime
beyond the validity of linearisation, which features underdamped relaxation oscillations of $\psi(t)$. 
Note that for the conservative case $\kappa=0$ this is replaced by undamped periodic oscillations~\cite{Altman2002}. 
At longer times, $t\kappa>1$, the oscillations are damped out but the amplitude $\psi(t)$ also decays to zero as $\exp(-\kappa t)$ due to the photon loss. However, the field rescaled by the occupation, $\bar{\psi}_i(t) = \psi_i(t)/\sqrt{n_i}(t)$ does reach a non-trivial steady state.  
This is shown in the top inset of Fig~\ref{fig:clean1}.  In contrast, if one considered
instead a case where $J/U < \left.J/U\right|_{\text{cr}}$ there would
be no instability. 
One should however note that while the distinction
  between initial stability and instability depends only on the
  parameters of the model, the asymptotic value of $\bar\psi(t)$ does
  depend (logarithmically) on the value of $\eta$ chosen --- for
  smaller $\eta$ the instability takes longer to reach the nonlinear
  regime, and so the average photon number at this point is smaller,
  affecting the final state reached. Further information about the
state can also be found from the correlation function $g_{2}(t) =
\textrm{Tr}[\hat{n}_i (\hat{n}_i-1) \rho_i(t)]/n_i(t)^2$. Again, because this quantity is normalized it asymptotically approaches a constant non-zero value (unless $n_0=1$). One may in fact show that if $J=0$, $g_2(t) = g_2(0) = 1-1/n_0$ remains fixed at the value determined by the initial Fock state.

\begin{figure}[h!]
  \begin{center}$
    \begin{array}{c}
      \includegraphics[width=0.45\textwidth]{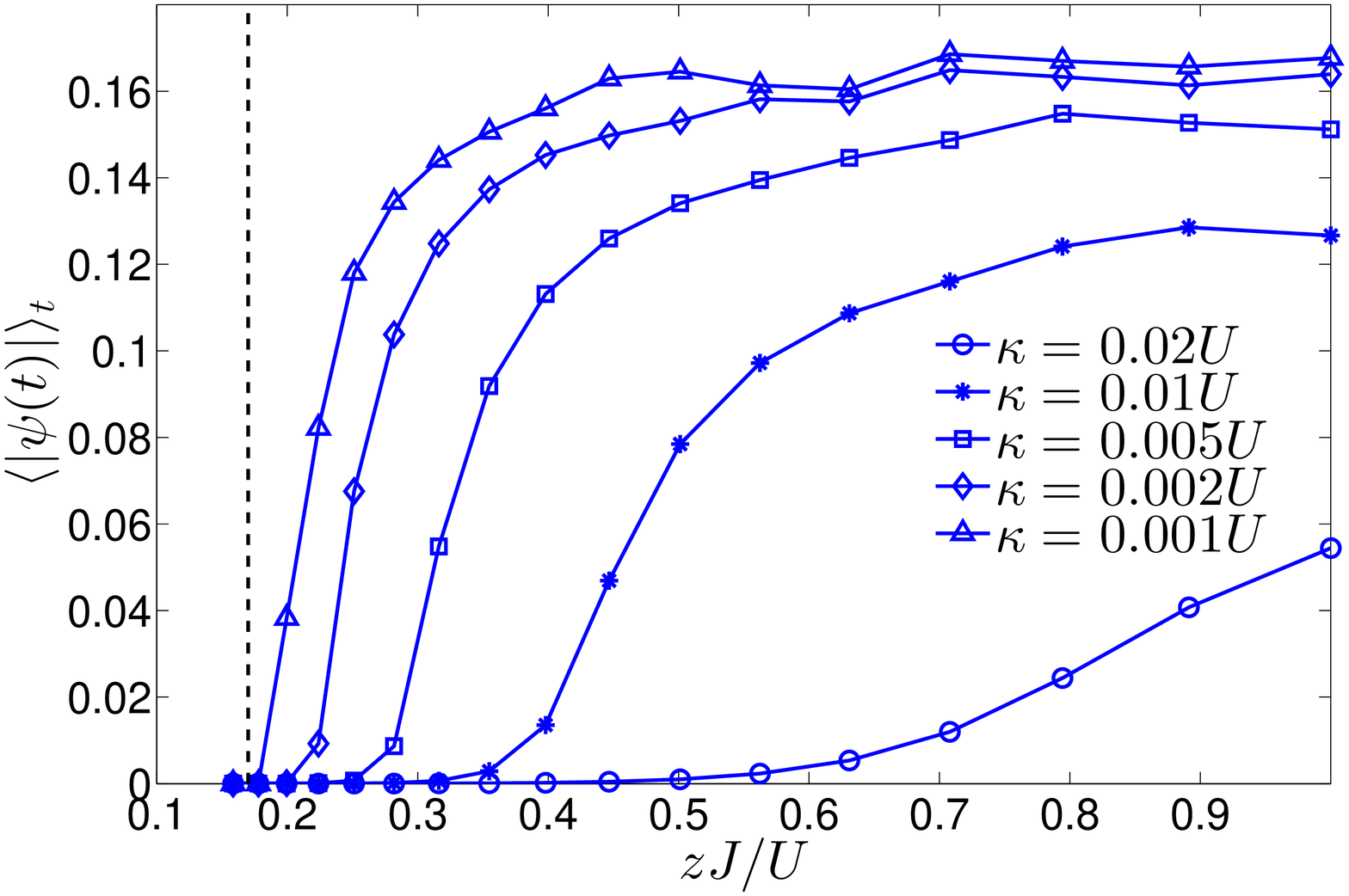} \\ 
      \includegraphics[width=0.45\textwidth]{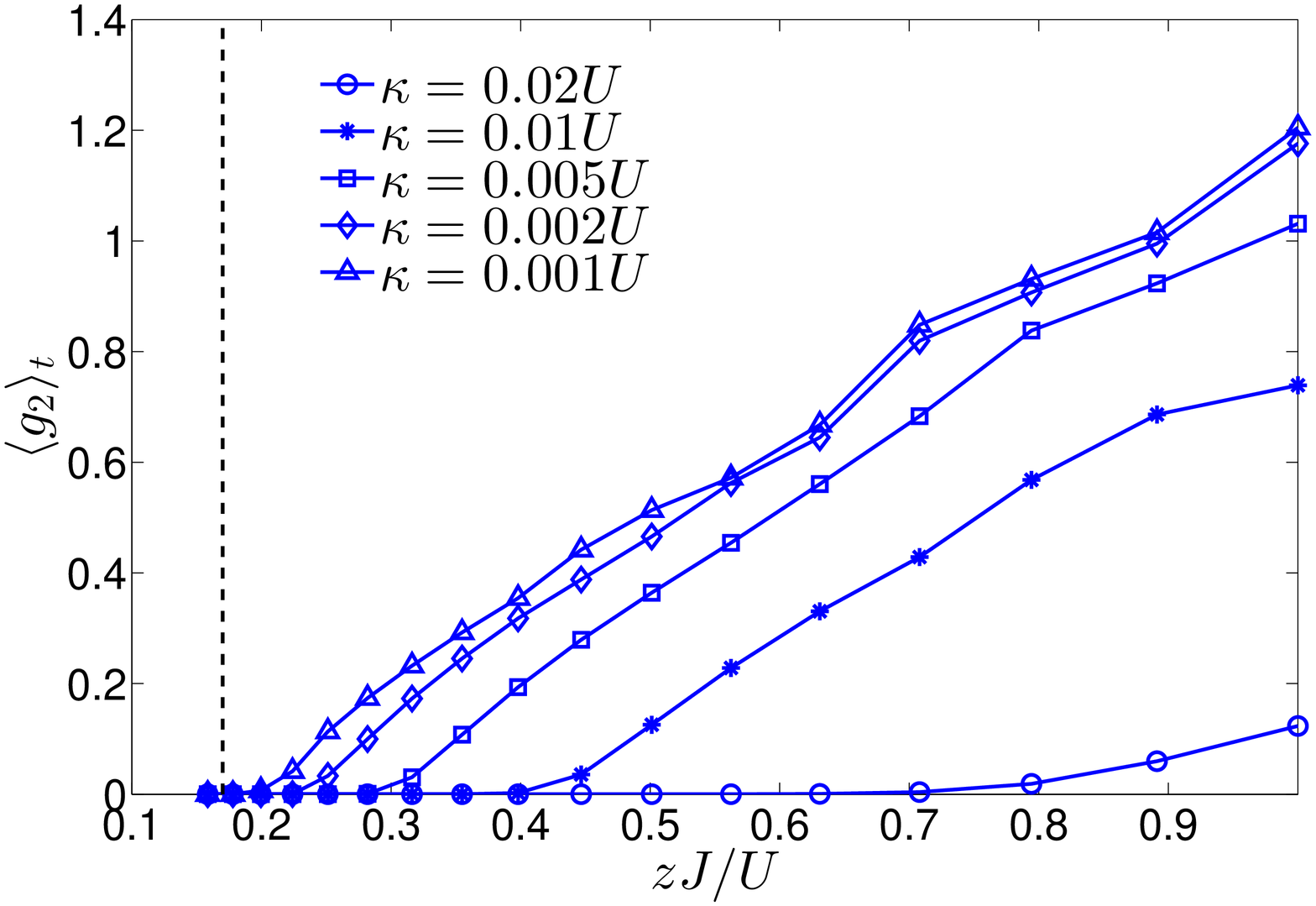}     
    \end{array}$
  \end{center}
  \caption{Time average of $|\psi(t)|$ (left panel) and $g_{2}(t)$
    (right panel) in the time interval $1<t\kappa<2$ as a function of
    the hopping amplitude $J/U$ for $\kappa=0.001U$ (triangles), 
    $\kappa=0.002U$ (diamonds),  $\kappa=0.005U$ (squares), $\kappa=0.01U$ (asterisks) and 
    $\kappa=0.02U$ (circles). The vertical dashed line denotes the 
    value at which the Mott insulator--superfluid transition occurs 
    in the equilibrium Bose-Hubbard model at integer filling ${n}_0=1$,
    $zJ/U|_{\textrm{cr}}\approx0.17$.}
  \label{fig:clean_int}
\end{figure}

Since the rescaled field $\bar{\psi}(t)$ and correlation function
$g_2(t)$ approach assymptotic values at late times, the behaviour for
a given set of initial conditions can be characterised by these
values.  Formally these are extracted by finding the time-averaged
values $\langle|\psi|\rangle_{t}$, $\langle g_{2}\rangle_{t}$,
averaging over a time window that neglects the intial transients as
proposed by~\citet{Tomadin2010}. Such an approach is illustrated in
Figure~\ref{fig:clean_int}, which shows the time-averaged $\langle
|\psi|\rangle_t$ and $\langle g_{2}\rangle_t$ in the time interval
$1<t\kappa<2$, as a function of the hopping amplitude $J/U$ and for 
five different values of the photon decay rate $\kappa$.  One may
clearly see that below a threshold value of $zJ/U$ both $\langle|\psi|\rangle_{t}$ and $\langle g_{2}\rangle_{t}$ vanish.  As
$\kappa \to 0$ this threshold approaches the equilibrium
superfluid--insulator transition which occurs at
$zJ/U|_{\textrm{cr}}\simeq0.17$ for ${n}_0=1$.  While these results
occur in mean-field theory, similar results have been reported by
using a cluster mean-field approach by~\citet{Tomadin2010}.

\subsection{Disordered case}
\label{sec:disordered-case}
\begin{figure}[ht!] 
  \begin{center}$
    \begin{array}{c}
      \includegraphics[width=0.45\textwidth]{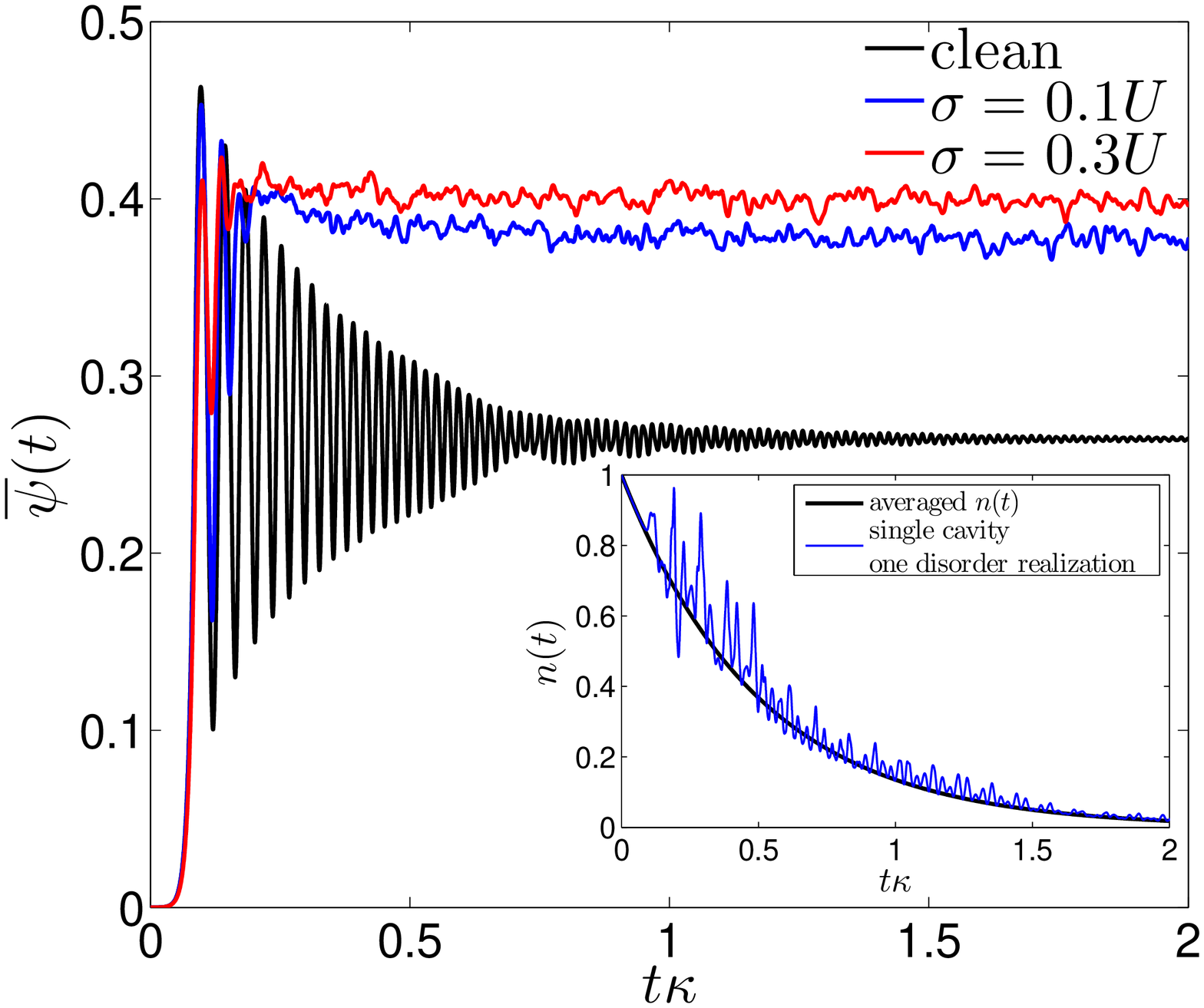}\\ 
      \includegraphics[width=0.45\textwidth]{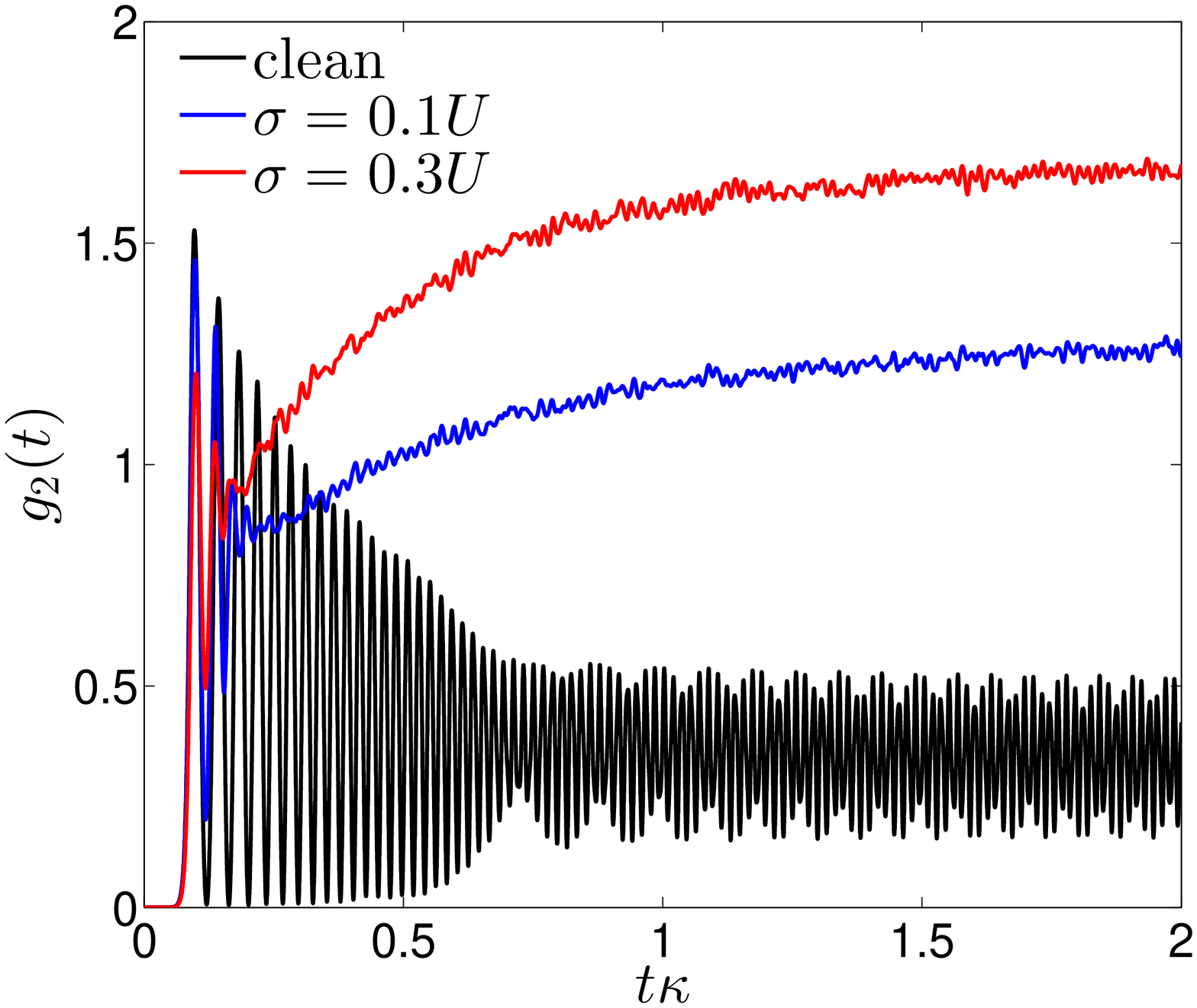}            
    \end{array}$
  \end{center}
\caption{Dynamics of a disordered one-dimensional array ($z=2$)
    consisting of $N=48$ coupled cavities in presence of a dissipation
    $\kappa=10^{-2}U$ and for $zJ=3U$. The simulation has been
    performed for 10 different realisations of Gaussian distributed
    cavity energies with $\sigma=0.1U$ (blue line) and $\sigma=0.3U$
    (red line). Top panel: Time evolution of the absolute value of the
    order parameter $|\psi(t)|$. The inset shows the evolution of the
    average filling $\langle n\rangle=n(t)$ for a single cavity and a
    single realisation of disorder with $\sigma=0.1U$ (blue line) and
    after averaging over all the cavities and all the realisations
    (black line). Bottom panel: Time evolution of the zero-time delay
    second order correlation function $g_{2}(t)$.}
  \label{fig:cleandis}
\end{figure}
\begin{figure}[ht!]
  \begin{center}$
    \begin{array}{c}
      \includegraphics[width=0.45\textwidth]{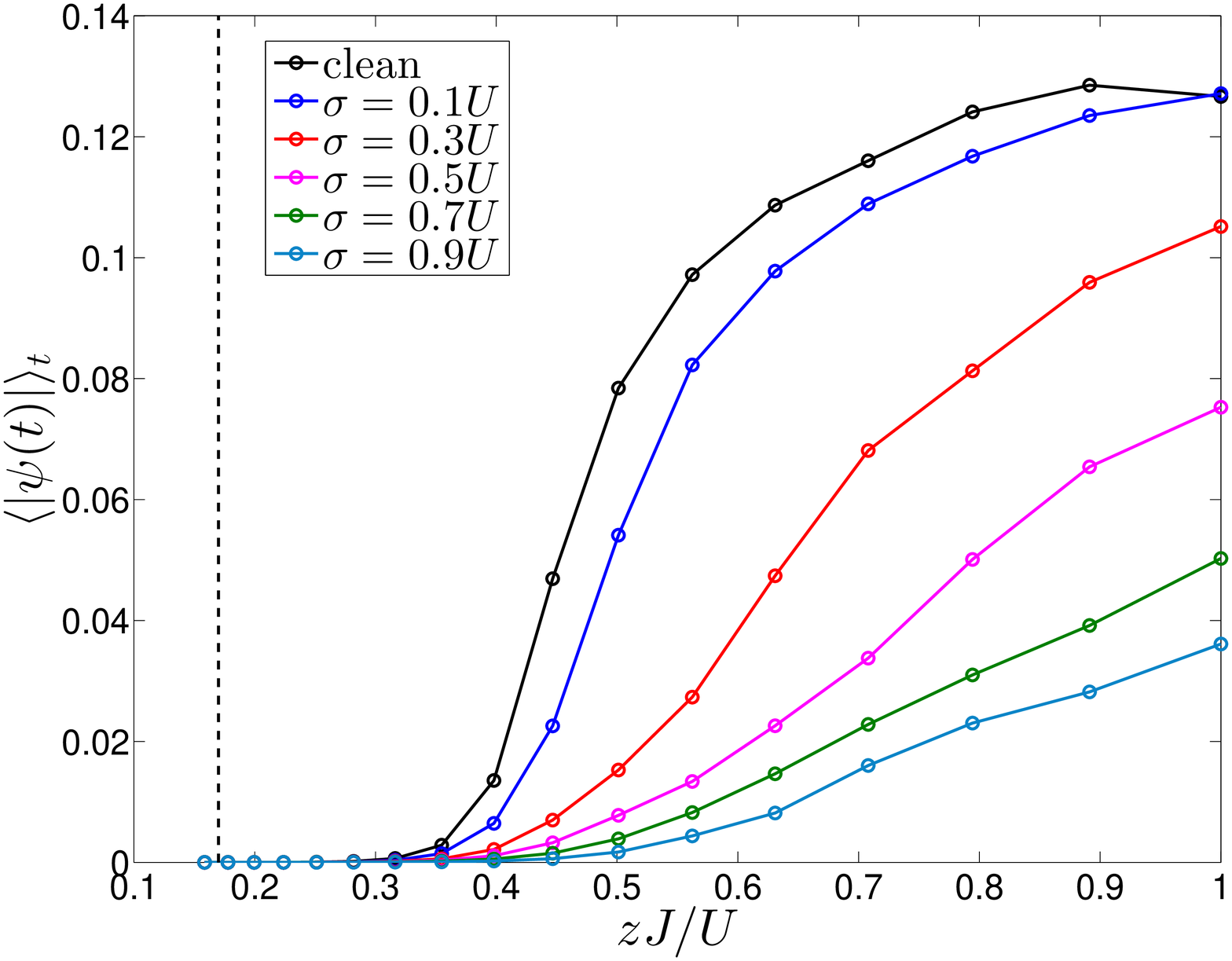} \\ 
      \includegraphics[width=0.45\textwidth]{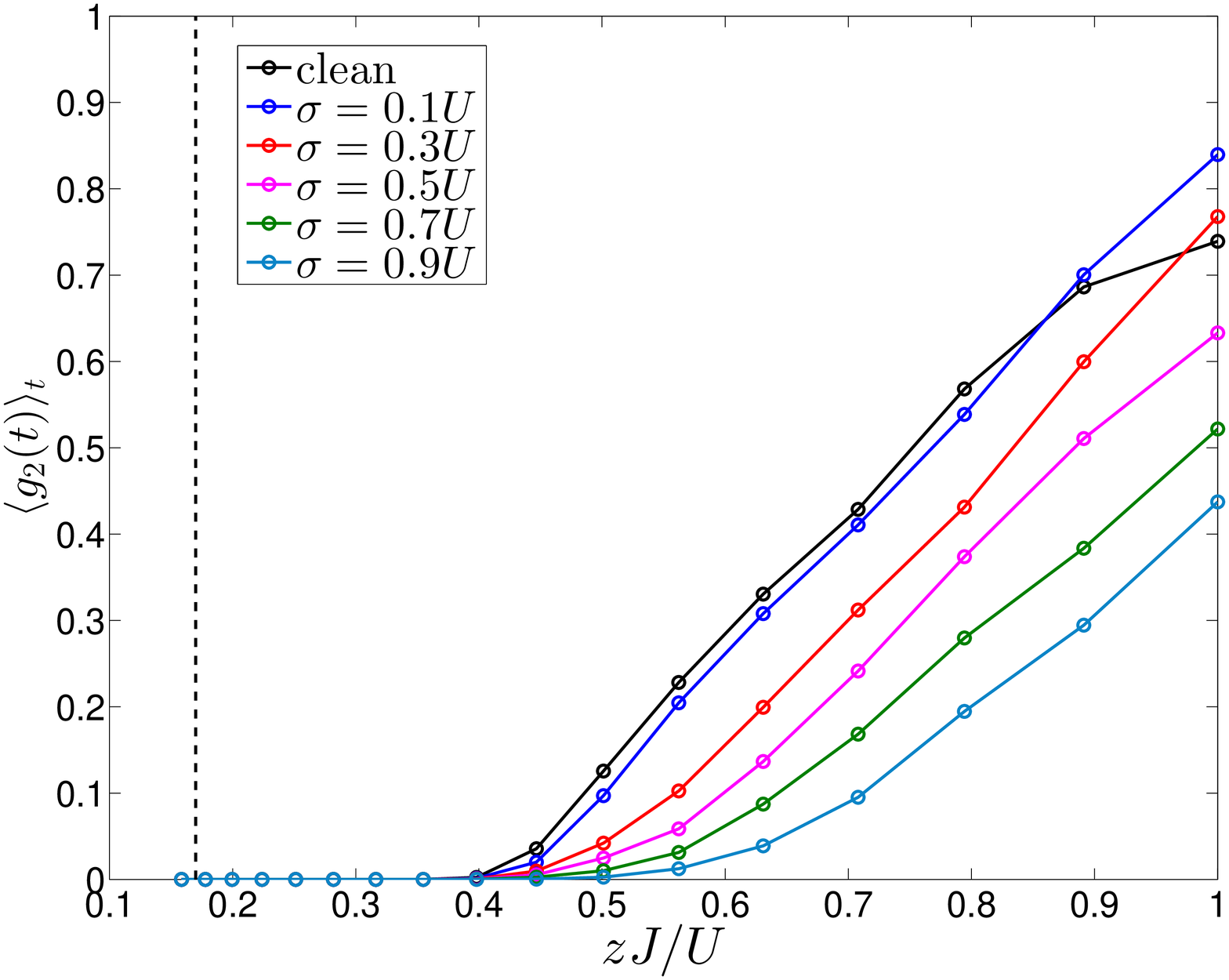}    
    \end{array}$
  \end{center}
  \caption{Time average of $|\psi(t)|$ (top panel) and $g_{2}(t)$
    (bottom panel) in the time interval $1<t\kappa<2$ as a function of
    the hopping amplitude $J/U$ for fixed decay $\kappa=0.01U$ and
    increasing disorder strength $\sigma$.  The vertical dashed line
    identifies the value at which the Mott insulator--superfluid
    transition occurs in the equilibrium Bose-Hubbard model at integer
    filling ${n}_0=1$, $zJ/U|_{\textrm{cr}}\approx0.17$.}
  \label{fig:dis_int}
\end{figure}
We now explore the role played by the on-site cavity disorder
$\eps_{i}$ in the non-equilibrium dynamics.  As discussed above, we
thus solve the Liouville problem Eq.~(\ref{eq:master}) drawing the
cavity energies $\eps_i$ from a Gaussian distribution with standard
deviation $\sigma$.  In order to characterise the properties of the
ensemble, rather than those specific to a particular realisation, we
average the expectations $|\psi|$ and $g_2$ over different
realisations of disorder, and additionally average over all sites
within a given realisation.

Figure~\ref{fig:cleandis} shows the time evolution of both the order
parameter (left panel) and the second-order correlation function
(right panel) for a 1D array consisting of $N=48$ cavities,
averaged over $10$ realisations of the energies.  Except
for the distribution of site energies $\eps_i$, all other parameters
are as in Fig.~\ref{fig:clean1}.  The site energies are drawn from
Gaussian distributions with $\sigma=0.1U$ (blue line) and $0.3U$ (red
line).  At short times $t\kappa<0.1$, the dynamics is characterised by
the same linear instability as is seen in the clean case (black line),
and both $\psi(t)$ and $g_2(t)$ increase exponentially.  At later
times however disorder strongly modifies the behaviour.  The
oscillations seen previously in the clean case are washed out by
averaging over different cavities, and so a quasi-steady state value
is reached earlier.
\begin{figure}[ht!]
\centering
      \includegraphics[width=0.48\textwidth]{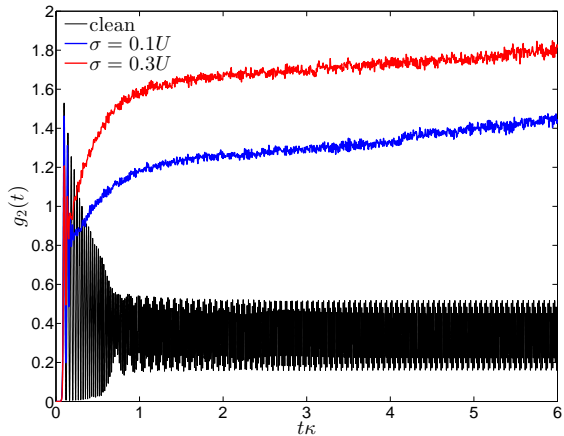} 
\caption{Dynamics of a disordered one-dimensional array ($z=2$) consisting of $N=48$ coupled cavities in presence of a dissipation 
$\kappa=10^{-2}U$ and for $zJ=3U$ 
The simulation has been performed for 10 different realisations of Gaussian distributed cavity energies with $\sigma=0.1U$ (blue line) and $\sigma=0.3U$ (red line).}
  \label{fig:g2_anomaly_zj3}
\end{figure}

As in the clean case, the appearance of a plateau at late times
suggests it is possible to characterise the evolution by its
asymptotic value (see discussion in
section~\ref{sec:rare-site-statistics}).  Figure~\ref{fig:dis_int}
shows the time-integrated $|\psi|$ (left panel) and $g_{2}$ (right
panel) as the ratio $J/U$ is varied at a fixed photon dissipation
constant $\kappa=10^{-2}U$ and for increasing values of disorder
strength. As in the clean case, a threshold value of $zJ/U$ is
required before the instability occurs.  
This threshold value of $zJ/U$ for the instability of the
$\psi=0$ state appears to increase with increasing disorder.  
In equilibrium there is a ``Bose glass'' phase between the Mott insulator
and the superfluid~\cite{Fisher1989}, where particles are no longer localised by
interactions, but are instead localised by disorder.  The critical
hopping $zJ/U$ for the equilibrium transition between the Bose glass
and superfluid phase increases with hopping, and so our observation of
increasing critical $zJ/U$.  The increasing critical $zJ/U$ we
observe for the threshold in the open system is consistent with this.

\section{Rare site statistics at late times}
\label{sec:rare-site-statistics}
\begin{figure}[h!]
  \begin{center}$
    \begin{array}{c}
      \includegraphics[width=0.45\textwidth]{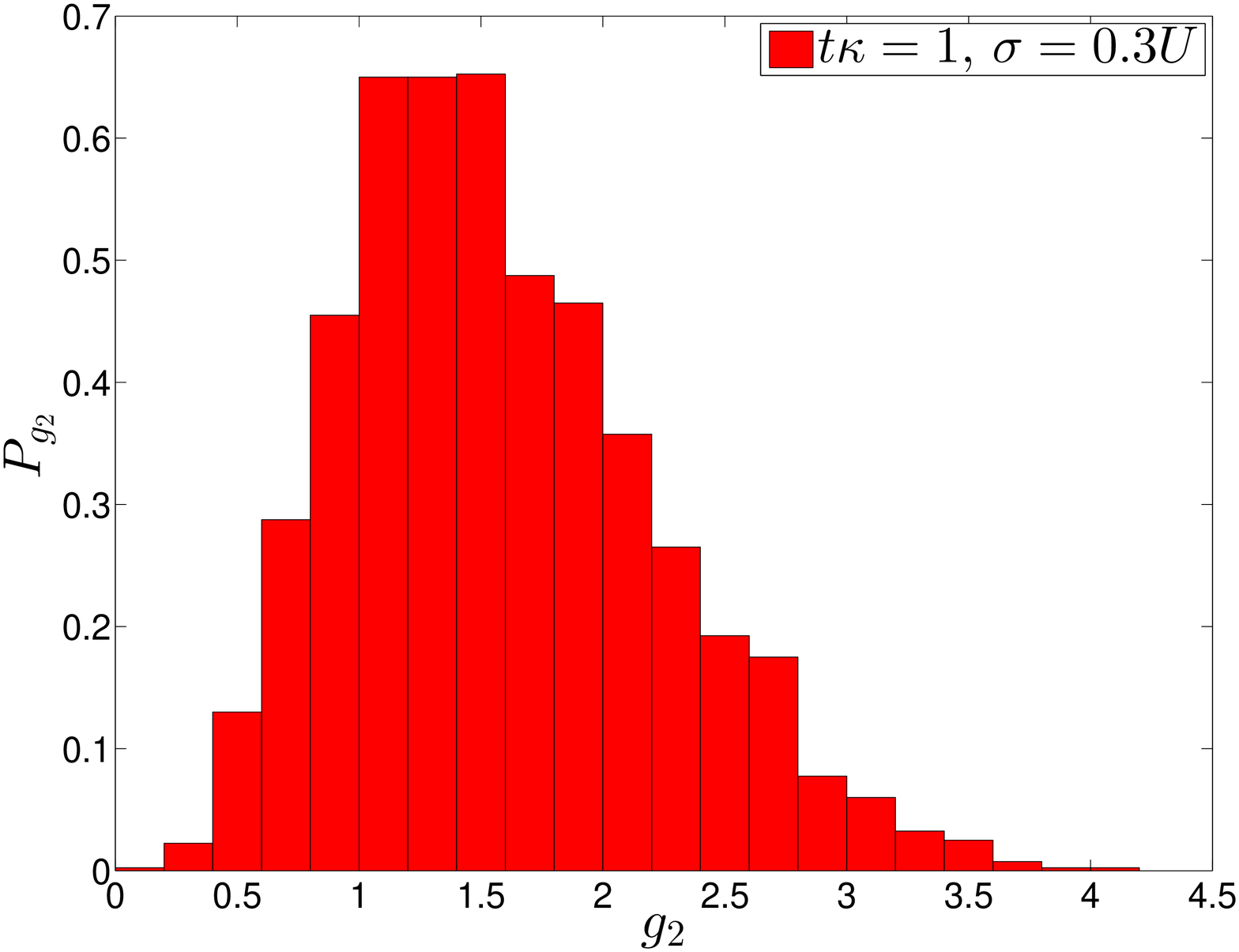} \\ 
      \includegraphics[width=0.45\textwidth]{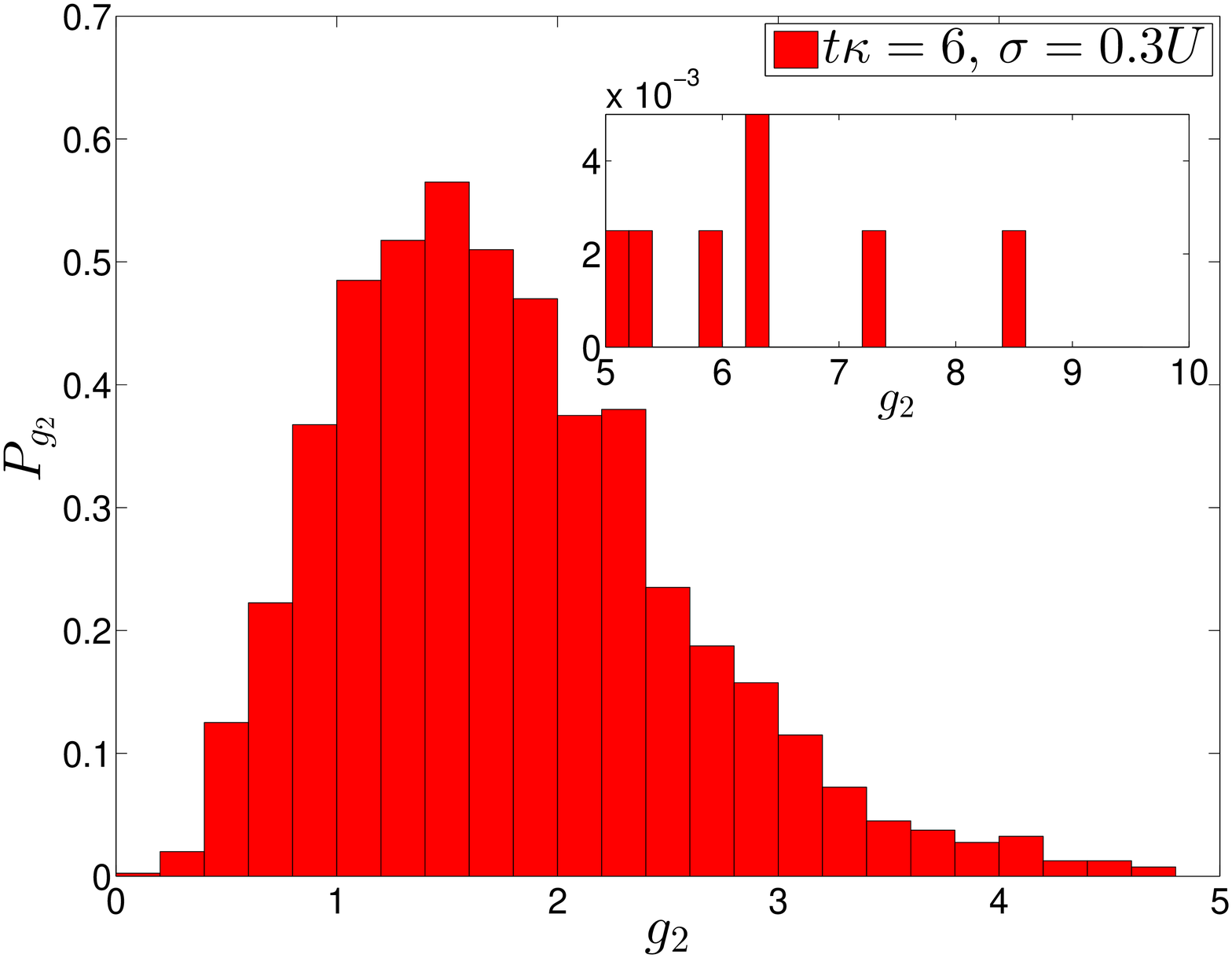}     
    \end{array}$
  \end{center}
  \caption{Probability density of $g_{2}$ calculated for a 1D array using $\kappa=0.01U $, $zJ=3U$ and $\sigma=0.3U$ (see the red curve in Fig. \ref{fig:g2_anomaly_zj3}). 
  The probability density has been evaluated using a sample of 2000 cavities, obtained simulating 20 different disorder realisations of an 
  array consisting of 100 cavities. Top panel: the probability density of $g_{2}$ at $t\kappa=1$;  Bottom panel: the probability density of 
  $g_{2}$ at $t\kappa=6$. The inset shows the occurrence of rare cavities with values of $g_{2}$ larger than 5.}
  \label{fig:g2_hist_zj3}
\end{figure}

\begin{figure}[t!]
\centering
      \includegraphics[width=0.45\textwidth]{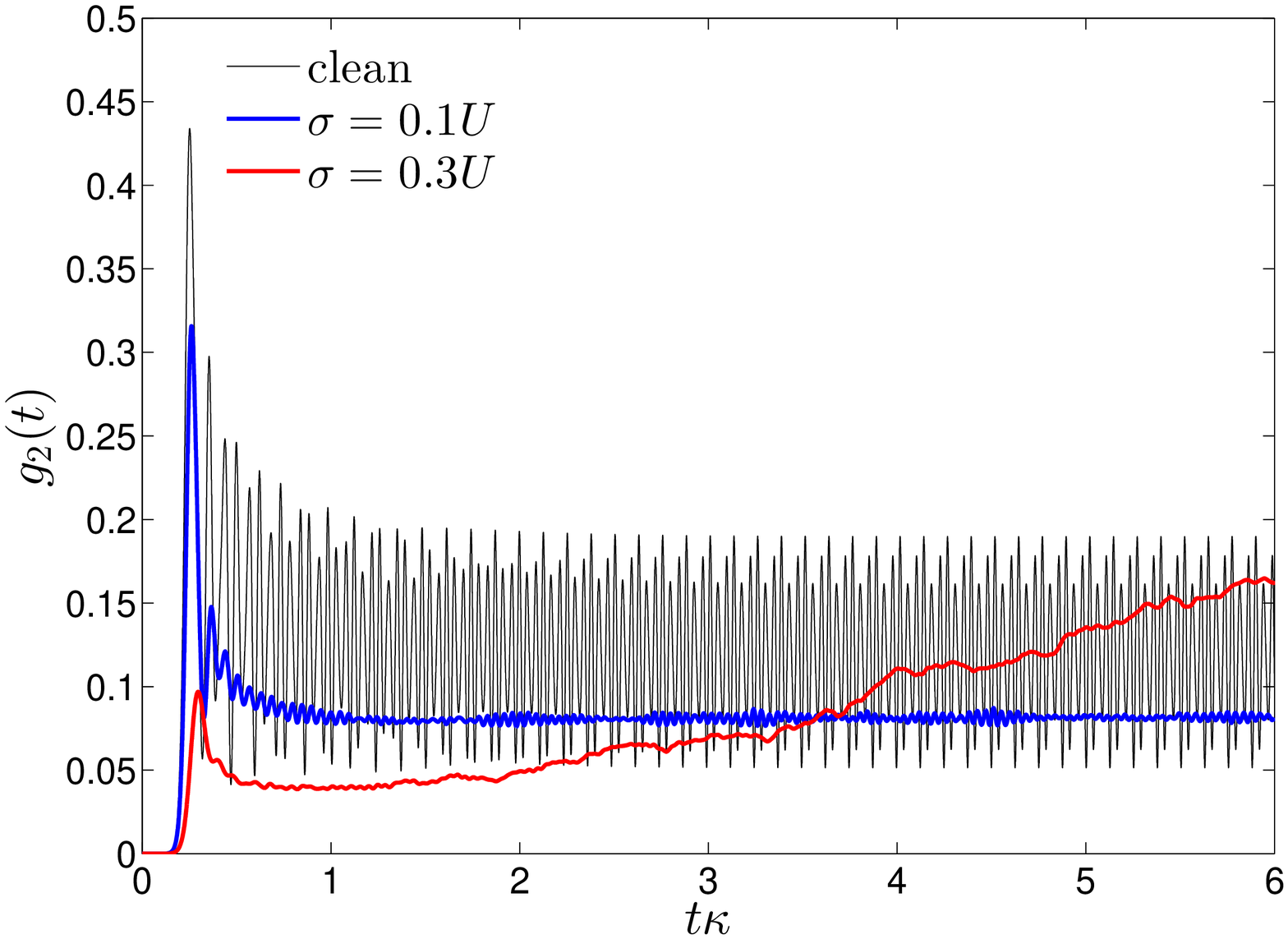} 
\caption{Dynamics of a disordered one-dimensional array ($z=2$) consisting of $N=48$ coupled cavities in presence of a dissipation 
$\kappa=10^{-2}U$ and for $zJ=0.5U$ 
The simulation has been performed for 10 different realisations of Gaussian distributed cavity energies with $\sigma=0.1U$ (blue line) and $\sigma=0.3U$ (red line).}
  \label{fig:g2_anomaly_zj05}
\end{figure}

\begin{figure}[t!]
  \begin{center}$
    \begin{array}{c}
      \includegraphics[width=0.45\textwidth]{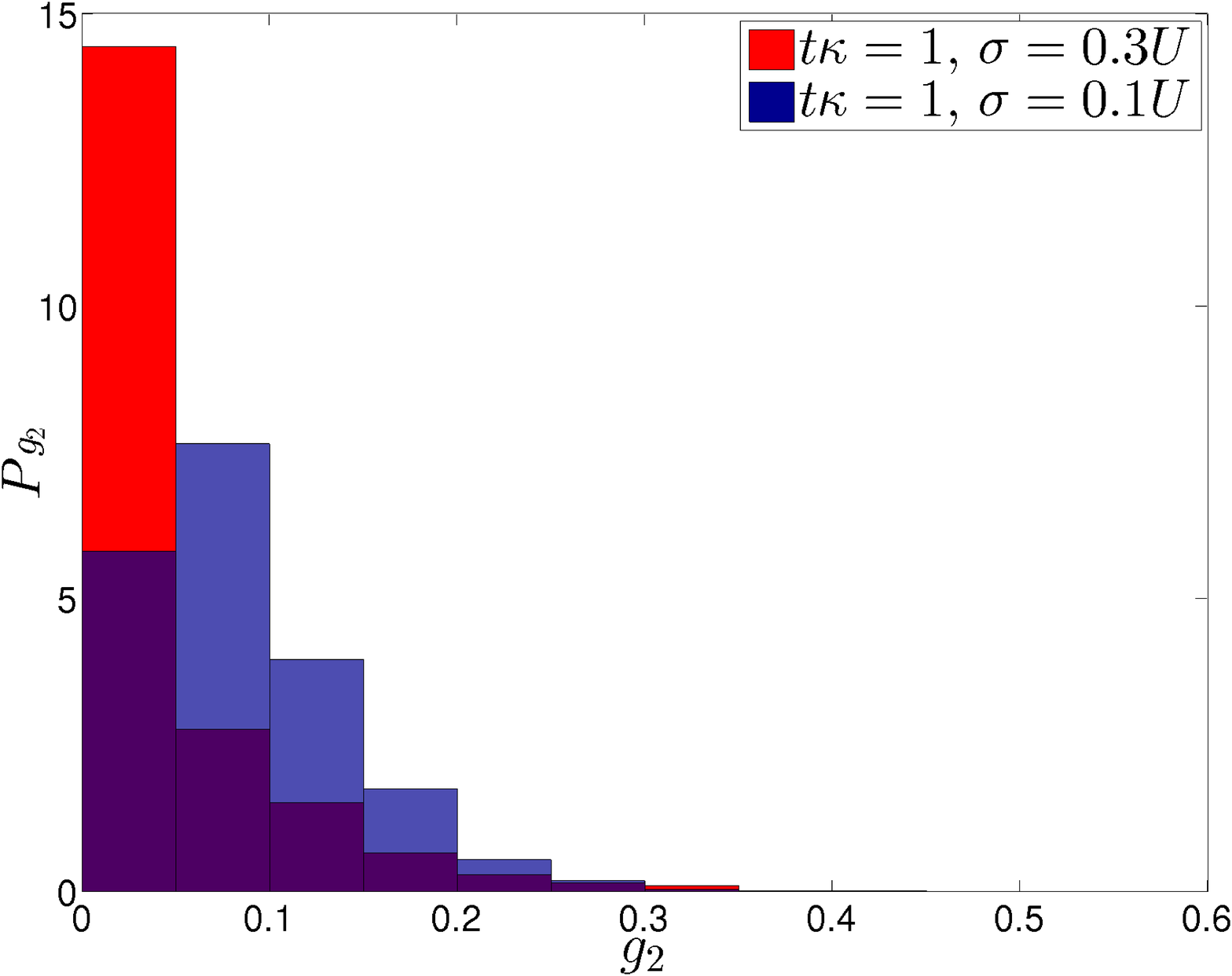} \\ 
      \includegraphics[width=0.45\textwidth]{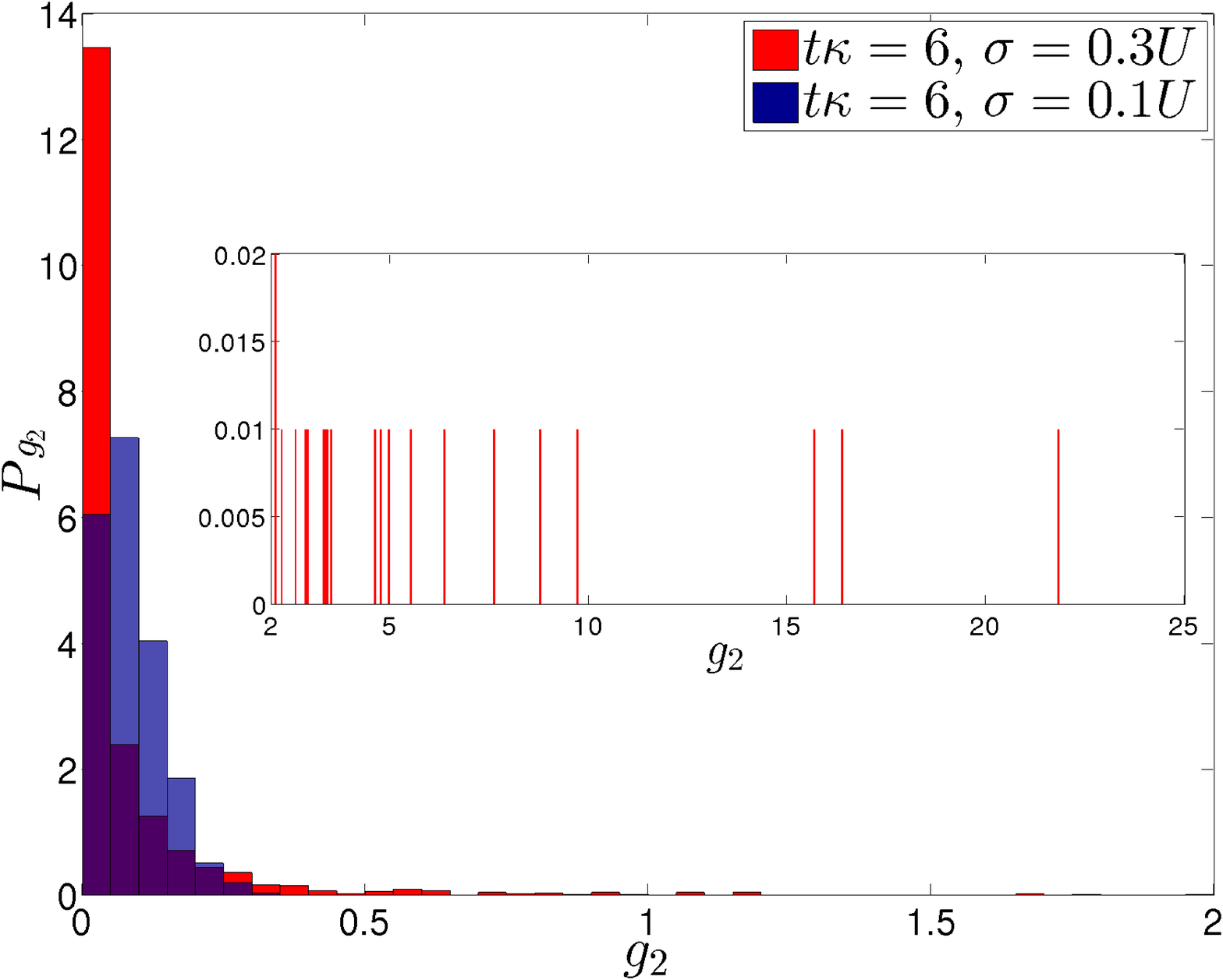}    
    \end{array}$
  \end{center}
  \caption{Probability density of $g_{2}$ calculated for a 1D array with $\kappa=0.01U $ and $zJ=0.5U$ as in Fig. \ref{fig:g2_anomaly_zj05}. 
  The probability density has been evaluated using a sample of 2000 cavities, obtained simulating 20 different disorder realisations of an 
  array consisting of 100 cavities. Top panel: the probability density of $g_{2}$ at $t\kappa=1$ for $\sigma=0.1U$ (blue) and $\sigma=0.3U$ (red); 
  Bottom panel: the probability density of $g_{2}$ at $t\kappa=6$. The inset shows that when $\sigma=0.3U$ some cavities exhibit exceptionally 
  large values of $g_{2}$ up to $\gtrsim\,20$.}
  \label{fig:g2_hist_zj05}
\end{figure}

While in the clean case, the plateau reached by $\bar{\psi}(t)$ and
$g_2(t)$ around $t\kappa \simeq 2$ reflects the asymptotic time
dependence, this turns out not to be the case for the disordered
lattice.  Despite the appearance of an apparent plateau seen in
figure~\ref{fig:cleandis} only indicates a temporary plateau.  At
later times, the values of $\langle g_2(t) \rangle_{\text{dis}}$
starts to rise further as shown in figure~\ref{fig:g2_anomaly_zj3}. 
Intriguingly, this rise of $\langle g_2(t) \rangle_{\text{dis}}$ at late times in fact reflects the existence of rare sites 
with large and exceptionally large values of $g_2(t)$ which dominate the disorder average. 
This is illustrated in figure~\ref{fig:g2_hist_zj3}, which shows the probability density of $g_{2}$ ($P_{g_{2}}$) calculated 
for the same hopping and decay constants used in figure~\ref{fig:g2_anomaly_zj3} using a sample of 2000 values of $g_{2}$,  
generated after the simulation of a 1D array of 100 cavities for 20 different realisations of disorder with $\sigma=0.3U$ (see the red curve in figure~\ref{fig:g2_anomaly_zj3}). The 
right panel shows that, at late times ($t\kappa=6$) some sites exhibit large values ($>\,5$) of $g_{2}$. 
This behaviour can be better understood by looking at figure~\ref{fig:g2_anomaly_zj05}, which illustrates for a smaller hopping $zJ=0.5U$ and  $\kappa=0.01U$, 
how the disorder average $\langle g_2(t) \rangle_{\text{dis}}$ rises when $\sigma=0.3U$ (see the red curve) as a result of the occurrence of rare sites with large values 
of $g_{2}$, as shown by the probability density evaluated at $t\kappa=1$ and $t\kappa=6$, shown in figure~\ref{fig:g2_hist_zj05}. It is worth noticing that all the anomalous sites 
(i.e. with $g_{2}\gg\,1$) are local minima of the potential (energy landscape), but not all the local minima exhibit such  anomalous values of $g_{2}$. Further, 
following the evolution to much later times becomes difficult, as the large ratio between different elements of the density matrix introduces numerical errors.

\section{Conclusions}
We have studied the time evolution of the disordered open Bose-Hubbard
model following an initially prepared Mott state.  As for the clean
case, a dynamical transition does occur between small and large values
of hopping, signalled by the asymptotic behaviour of the rescaled
field $\bar{\psi}(t)$.  We find that in line with the equilibrium
expectations, the presence of disorder does increase
$\left.zJ/U\right|_{\text{cr}}$, i.e.  does increase the hopping
required for the superfluid instability to develop.  However, even
within mean-field theory, disorder can produce anomalous long time
dynamics, where ensemble averages are dominated by the effect of rare
sites.

\section{Acknowledgements}
We acknowledge financial support from an ICAM-I2CAM postdoctoral fellowship 
for the execution of this work. C.C. gratefully acknowledges support from ESF (Intelbiomat
Programme), R.F. from IP-SIQS and PRIN (Project 2010LLKJBX), J.K. from EPSRC
programme ``TOPNES'' (EP/I031014/1) and EPSRC grant (EP/G004714/2), and H.E.T. from NSF CAREER
(DMR-1151810) and The Eric and Wendy Schmidt Transformative Technology Fund. We
acknowledge helpful discussions with Peter Littlewood.

\newpage
%

\end{document}